\documentclass[conference]{IEEEtran}
\IEEEoverridecommandlockouts
\usepackage{cite}
\usepackage{epstopdf}
\usepackage{amsmath,amssymb,amsthm,mathrsfs,amsfonts,dsfont}
\usepackage{algorithmic}
\usepackage{graphicx}
\usepackage{color}
\usepackage{textcomp}

\newcommand{\site}{\boldsymbol{s}}
\newcommand{\cluster}{\boldsymbol{c}}
\def\BibTeX{{\rm B\kern-.05em{\sc i\kern-.025em b}\kern-.08em
    T\kern-.1667em\lower.7ex\hbox{E}\kern-.125emX}}
\begin{document}

\title{Interference Analysis in Dynamic TDD System Combined or not With Cell Clustering Scheme}

\author{\IEEEauthorblockN{Jalal Rachad$~^{*+}$, Ridha Nasri$~^*$, Laurent Decreusefond$~^+$}
	\IEEEauthorblockA{$~^*$Orange Labs: Direction of Green transformation, Data knowledge, traffic and resources Modelling,\\ 40-48 avenue de la Republique
		92320 Chatillon, France\\ 
		$~^+$LTCI, Telecom ParisTech, Universite Paris-Saclay,\\
		23 avenue d'Italie 75013, Paris, France\\
		Email:\{$^{*}$jalal.rachad,  $~^*$ridha.nasri\}@orange.com, $^+$Laurent.Decreusefond@mines-telecom.fr}}

\maketitle
\begin{abstract}
Dynamic Time Division Duplex (TDD) has been introduced as a solution to deal with the uplink and downlink traffic asymmetry, mainly observed for dense heterogeneous network deployments. However, the use of this feature requires new interference mitigation schemes capable to handle two additional types of interferences between cells in opposite transmission cycle: downlink to uplink and uplink to downlink interferences. Among them, Cell clustering has been proposed as an efficient solution to minimize inter-cell interferences in opposite transmission directions and somehow responds to the requirements of enhanced Interference Mitigation and Traffic Adaptation (eIMTA) problem. This work is devoted to provide a new analytical approach to model inter-cell interferences and quantify performances of Dynamic TDD system in terms of SINR (Signal to Interferences plus Noise Ratio) distribution. Analytical system performance investigation concerns two scenarios: \textit{i}) basic Dynamic TDD without any other feature and \textit{ii}) Dynamic TDD with interference mitigation schemes.  
\end{abstract}


\section{Introduction}
Future mobile networks are expected to support the proliferation of numerous real-time applications requiring high data rates, significant traffic variations and low latency. Dynamic Time Division Duplex (D-TDD) has been proposed in order to deal with traffic asymmetry since it enables the dynamic adjustment of UL and DL resource transmissions according to the traffic variations. However, D-TDD system is severely limited by a strong mutual interference between the UpLink (UL) and DownLink (DL) transmissions because those two directions share the same frequency band. Hence, two types of interference appear: DL to UL (impact of DL other cell interferences on UL signal received by the studied cell) and UL to DL (impact of UL mobile users transmission, located in other cells, on DL signal received by a mobile user located in the studied cell). Those additional interferences, mainly DL to UL, are usually more difficult to deal with because of the LOS (Line Of Sight) between highly elevated base stations transmitting with high power level and also because the mobiles can move around randomly. In order to mitigate interferences in D-TDD systems, 3GPP (3rd Generation Partnership Project) standard advices new approach for enhanced Interference Mitigation and Traffic Adaptation (eIMTA) in dynamic environment \cite{astely2013lte}. A cell clustering scheme can be used so that cells suffering from mutual high DL to UL interferences can be gathered in the same cluster and use the same UL-DL configurations.

The available scientific literature of interference investigation in dynamic TDD system is quite rich. The first study dates back to 2002 with the work in \cite{jeong2002cochannel} where performances of a D-TDD fixed cellular network in UL transmission were investigated. Authors in \cite{jeong2002cochannel} proposed a time slot assignment method to improve the UL outage performances. Performances of D-TDD system were also provided in \cite{yu2015dynamic} for a particular small cells' architecture known as phantom cells in UL and DL transmission directions. For the analytical approach, \cite{yu2015dynamic} used tools from stochastic geometry to model phantom cells and user locations in order to derive SINR distributions. Likewise, performances of D-TDD enabled mmWave cellular networks were discussed in \cite{kulkarni2017performance}. Additionally, in order to make D-TDD feasible, some interference mitigation techniques have been proposed in literature, such as cell clustering \cite{gao2015performance}, \cite{lin2015dynamic}. It was discussed in \cite{lin2015dynamic} a soft reconfiguration method based on cell clustering so as to allow cells in the same cluster to change dynamically the UL/DL configuration but inter-cluster interference still exists.

Always in the same context, we propose in this paper a new approach of performance investigation in D-TDD including an interference mitigation scheme based on  cell clustering. The analytical approach of the paper adopts a hexagonal geometry of macro cells and a spatial random distribution of small cells. The paper contributions cover in particular the explicit evaluation of $ISR$ (Interferences over Signal Ratio) in each position of the network and the distribution of $SINR$ (Signal to Interferences plus Noise Ratio) for a typical cell. The first metric is useful for link budget tool in which the expression of the average perceived interference is required in each position, whereas the second metric is directly related to throughput distribution, so it is useful for cell throughput dimensioning in D-TDD systems.\\

The rest of this paper is organized as follows: In Section II, system models, notations and assumptions are provided. Different scenarios of interferences, introduced by D-TDD concept, are also highlighted. In section III, we provide analytical results regarding $ISR$ expressions for UL and DL under uniform user locations distribution. In Section IV, DL to UL interference in a heterogeneous deployment is investigated considering cell clustering scheme. Section V concludes the paper.

\section{System Models and Notations}

We consider a hexagonal cellular network denoted by $\Lambda$ with an infinite number of macrocells having an intersite distance between them denoted by $\delta$. The hexagonal model means that for each node $\site$ $\in$ $\Lambda$, there exists a unique (u,v) $\in$ $\mathbb{Z}^2$ such that $\site = \delta( u + ve^{i\frac{\pi}{3}})$, we denote by $\site_0$ the name of the serving cell located at the origin of $\mathbb{R}^2$. Antenna in each site is assumed to have an omni-directional radiation pattern and covers a geographical area named Voronoi cell, having a cell radius denoted by R. Furthermore, the location of a mobile served by $\site_{0}$ is denoted by $m$ such that $m = re^{i\theta}$ where (r, $\theta$) are the polar coordinates in the complex plane. We denote also by $n$ the geographical location of a mobile served by a cell $\site$ $\in$ $\Lambda^{*}$ in the plane, where $\Lambda^{*}$ is the lattice $\Lambda$ without the serving cell $\site_0$. Location $n$ is written in the complex plane by $n = \site + \rho e^{i\phi}$, where $\rho$ and $\phi$ represents respectively the distance and the angle between $n$ and $\site$. Furthermore, it is assumed that the locations of mobile $n$ in the plane are uniformly distributed.\\
The  propagation loss undergone by the signal transmitted in downlink by cell $\site$ $\in$ $\Lambda$ to location $m$ is modeled by
\begin{equation}
L(s,m)=\left|s-m\right|^{2b}\nonumber
\label{pathloss}
\end{equation}
where $2b>2$ is the path loss exponent.\\
In uplink transmission, power control is applied to PUSCH (Physical Uplink Shared CHannel) channel in order to set the required mobile transmitted power. In this paper, it is modeled by the fractional power control model, i.e., the path loss is partially compensated by the power control \cite{castellanos2008performance}. The transmitted power by the mobile location $n$ to its serving cell $\site$ is then written
\begin{equation}
P(n,\site)= P^*(\site) \left(\left|n - \site \right|^{2b}\right)^{k}
\label{power control}
\end{equation}
where $P^*$($\site$) is the target power cell specific and $k$ $\in$ [0 1] is the power control compensation factor. When $k=1$ the power control scheme totally indemnifies the path loss in order to reach the target power $P^*(\site)$. For the case $0<k<1$ the path loss is partially compensated and mobile users in cell edge create less interferences because their transmitted power is reduced. Without loss of generality, we consider that $P^*(\site)$ is the same for all cells and we denote it by $P^*$. We assume also that all cells transmit in downlink with the same power level $P$. Power values $P$ and $P^*$ are supposed to include the path loss constants and antenna gains of base station and user equipment.\\
\begin{figure}[htbp]
	\centering
	\includegraphics[height=5cm,width=8cm]{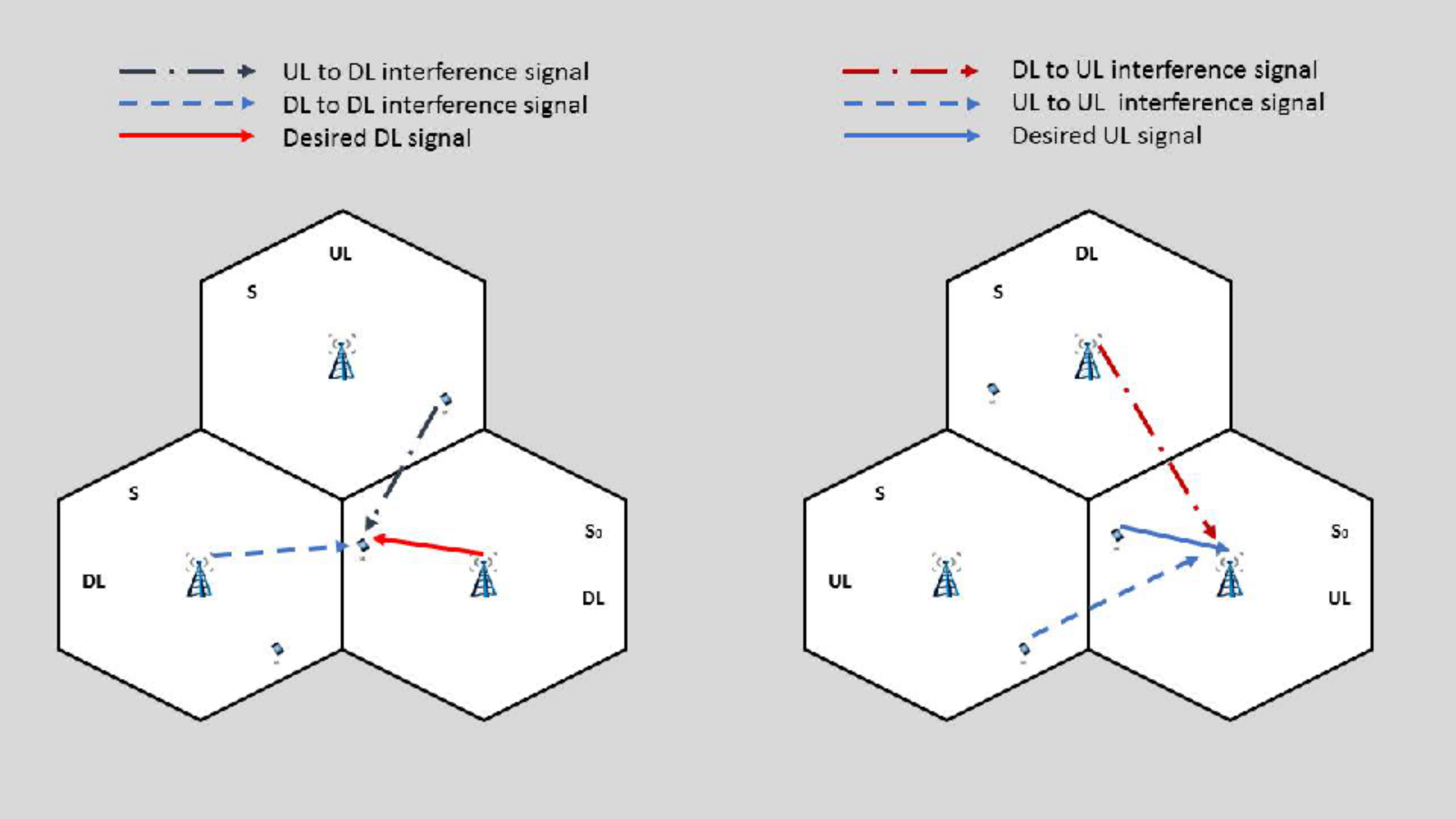}
	\caption{D-TDD UL to DL interference (left) and DL to UL interference (right)}
	\label{interfer}
\end{figure}

To model the D-TDD system, we assume that all cells initially operate synchronously in DL or UL. This setup can be considered as a baseline scenario characterizing performances of existing synchronous TDD system e.g., TD-LTE. After a period, it is assumed that all cells randomly select uplink or downlink transmission direction based on traffic conditions. Four types of interferences henceforth appear depending on the transmission direction: \textit{i}) when $\site_{0}$ transmits to a given mobile location, DL and UL interferences' effect on DL useful transmission appears; \textit{ii}) when $\site_{0}$ receives signals from mobiles,  UL and DL interferences' impact on UL transmission rises (Fig. \ref{interfer}). It is considered hereafter that the scheduler does not allocate the same spectral resources to different mobile users in one cell at the same time (e.g., TD-LTE scheduling). So, intra-cell interferences is not considered.

To model interferences in D-TDD system, we denote by $\alpha_d$ and $\alpha_u$ the percentage of cells that are respectively in DL and UL transmission, such that $\alpha_d + \alpha_u \leq$ 1. Hence, the Interference to Signal Ratio ($ISR$) experienced in DL transmission by a mobile location $m$ connected to cell $\site_0$ is 
\begin{equation}
D(m)= \alpha_{d}~D_{\downarrow}(m) + \alpha_{u}~D_{\uparrow}(m)
\label{ISRtotalDL}
\end{equation}
with $D_{\downarrow}$ and $D_{\uparrow}$ are respectively DL to DL and UL to DL interferences experienced during the DL cycle.\\
Likewise, the $ISR$ experienced by cell $\site_0$ in UL transmission cycle is
\begin{equation}
U(m)= \alpha_{u}~U_{\uparrow}(m) + \alpha_{d}~U_{\downarrow}(m)
\label{ISRtotalUL}
\end{equation}
 where $U_{\uparrow}$ and $U_{\downarrow}$ are respectively UL to UL and DL to UL interferences experienced during the UL cycle.

\section{Dynamic TDD Interference derivation }

\subsection{Downlink $ISR$ derivation $D(m)$}
\subsubsection{Expression of DL to DL $ISR$ $D_{\downarrow}(m)$}

In \cite{nasri2016Analytical} authors showed that the downlink $ISR$ function of a location $m = re^{i\theta}$ in hexagonal cellular network with infinite number of cells admits a series expansion on $r$ and $\theta$ and is a very slowly varying function on $\theta$. Taking $x=\frac{r}{\delta}$ such that $x<1$ ($x<1/\sqrt{3}$ condition always satisfied in hexagonal network), the expression of $D_{\downarrow}$ is recalled from \cite{nasri2016Analytical}
\begin{equation}
D_{\downarrow}(m) = \frac {6x^{2b}}{\Gamma(b)^{2}}\sum_{h=0}^{+\infty}\frac{\Gamma(b+h)^{2}}{\Gamma(h+1)^{2}}\omega(b+h)x^{2h}
\label{ISRdlm}
\end{equation}
where $\Gamma(.)$ is the Euler Gamma function  and
\begin{equation}
\omega(z) = 3^{-z}\zeta(z)\left( \zeta(z,\frac{1}{3})-\zeta(z,\frac{2}{3})\right),
\label{omega}
\end{equation}
with $\zeta(.)$ and $\zeta(.,.)$ are respectively the Riemann Zeta and Hurwitz Riemann Zeta functions \cite{abramowitz1964handbook}.

\subsubsection{Expression of UL to DL $ISR$ $D_{\uparrow}(m)$}

The UL to DL interferences is generated from mobile users located at other cells, mainly from those located at the border of cells adjacent to cell $\site_0$. Since there is only one mobile user transmitting at the same time in UL for each cell, the total UL to DL $ISR$ can be evaluated by averaging over locations $n\in\site$ and then summing over $\site\in$ $\Lambda^{*}$. So, if we assume that location $n$ is uniformly distributed in cell $\site$, $D_{\uparrow}(m)$ is mathematically written as
	\begin{equation}
	D_{\uparrow}(m)= \frac{1}{\pi R^{2}}\int \limits_{0}^{R} \int \limits_{0}^{2 \pi}\sum_{\site \in \Lambda^{*}}\frac {P^{*}~\rho^{2bk}~r^{2b}}{P\left|\site + \rho e^{i\phi}- r e^{i\theta} \right|^{2b}}\rho d\rho d\varphi
	\label{ISRultodl}
	\end{equation}
To evaluate equation (\ref{ISRultodl}), we can proceed analogously to the proof of $ISR$ formulas in hexagonal omni-directional networks provided in \cite{nasri2016Analytical}. We start by taking $m^{'}= re^{i\theta} - \rho e^{i\phi}$. It is obvious that $\left|m^{'}\right| < \left|\site \right|$. It follows from \cite{nasri2016Analytical} that the sum over $\site$ inside the double integral admits a series expansion on $\left|m^{'}\right|/\delta$ as in (\ref{ISRdlm}). Using formula (\ref{ISRdlm}) and writing $\left|m^{'}\right|$ in terms of $r$, $\theta$, $\rho$ and $\phi$, (\ref{ISRultodl}) becomes
{
	
	\begin{align}
	D_{\uparrow}(m) &=\frac{6 P^{*}x^{2b}}{P \pi R^{2}~\Gamma(b)^{2}}\int \limits_{0}^{R} \int \limits_{0}^{2 \pi}\sum_{h=0}^{+\infty} \frac{\Gamma(b+h)^{2} \omega(b+h)}{\Gamma(1+h)^{2} \delta^{2h}} \times\nonumber\\
	&(r^{2} + \rho^{2} )^{h} (1- \frac{2r \rho}{r^{2} + \rho^{2}} \cos ( \phi ))^{h} \rho^{2bk+1} d\rho d\phi 
	\label{ISRultodl3}
	\end{align}
}
The sum and integrals of (\ref{ISRultodl3}) can be switched and the inside integral can be evaluated by expanding $(1- \frac{2r \rho}{r^{2} + \rho^{2}} \cos ( \phi ))^{h}$ as a binomial sum. After few derivations of known special integrals and simplifications, the UL to DL $ISR$ $D_{\uparrow}(m)$ can be evaluated by the following convergent series on $x=r/\delta$
{	
	\begin{align}
	D_{\uparrow}(m) &= \frac{6 P^{*} x^{2b} R^{2bk}}{P~\Gamma(b)^{2}}\sum_{h=0}^{+\infty}\sum_{n=0}^{\lfloor \frac{h}{2} \rfloor}\sum_{i=0}^{h-2n} \frac{\Gamma(b+h)^{2} \omega(b+h)}{\Gamma(n+1)^{2}\Gamma(h+1)}\times\nonumber\\
	&\frac{ (\frac{R}{\delta})^{2n+2i}~x^{2h-2n-2i}}{\Gamma(i+1)\Gamma(h-2n-i+1) (n+i+bk+1)}
	\label{ultodlisrfinal}
	\end{align}
}
Since $x < 1/\sqrt{3}$ for hexagonal model, it is obvious that the first elements of this series are sufficient to numerically evaluate $D_{\uparrow}$.
%
%

\subsection{Uplink $ISR$ derivation $U(m)$}

In this part, we will derive the analytical expression of the UL interference to signal ratio. The UL signal received from location $m$ at cell $\site_0$ experiences interferences coming from cells transmitting in DL and also from mobiles in adjacent cells which are in UL transmission cycle. The following results may be proved in much the same way as $D_{\downarrow}$ and $ D_{\uparrow}$ in the previous section.

\subsubsection{UL to UL $ISR$ $U_{\uparrow}(m)$}

The UL interference is generated by mobiles in neighboring cells which are randomly distributed in the network as opposed to the DL direction where cells' positions are fixed. Thus recalling the fact that mobile location $n$ is uniformly distributed in cell $\site$ and taking into account the definition of the transmitted power with fractional power control model given by equation (\ref{power control}), $U_{\uparrow}(m)$ can be expressed as
{\begin{align}
	U_{\uparrow}(m) =& \frac{1}{\pi R^2}\int \limits_{0}^{R} \int \limits_{0}^{2 \pi} \sum_{\site \in \Lambda^{*}} \frac{\rho^{2bk}~\left|\site+\rho~e^{i\phi}\right|^{-2b}}{r^{2b(k-1)}} \rho d\rho d\phi \nonumber\\
	=& A_{1}(b)~x^{2b(1-k)}
	\label{fds}
	\end{align}
}
where
\begin{equation}
A_{1}(b)=\frac{6(R/\delta)^{2bk}}{\Gamma(b)^{2}} \sum_{h=0}^{+\infty}\frac{\Gamma(b+h)^{2}~\omega(b+h)}{\Gamma(h+1)^{2}~(bk+h+1)}(R/\delta)^{2h}\nonumber
\label{ddulinterference}
\end{equation}

\subsubsection{DL to UL $ISR$ $U_{\downarrow}(m)$}

The signal coming from neighboring cells is often very strong with respect to mobile transmit power, especially if neighboring cells' antennas are in LOS condition or inter-site distance is lower (path loss is low). Contrary to the UL to UL interference, here the interfering signals come from cells, which have fixed positions. Hence, under the same system model assumptions, $U_{\downarrow}$ is given by

\begin{equation}
U_{\downarrow}(m) = \sum_{\site \in \Lambda^{*}} \frac{P \left|\site\right|^{-2b}}{P^{*}~r^{2b(k-1)}} = A_{2}(b)x^{2b(1-k)}
\label{dltoulinterference}
\end{equation}
where $A_{2}(b)=\frac{P~\omega(b)}{P^{*}~\delta^{2bk}}$.\\

\begin{figure}[htbp]
	\centering
	\includegraphics[height=5.5cm,width=8cm]{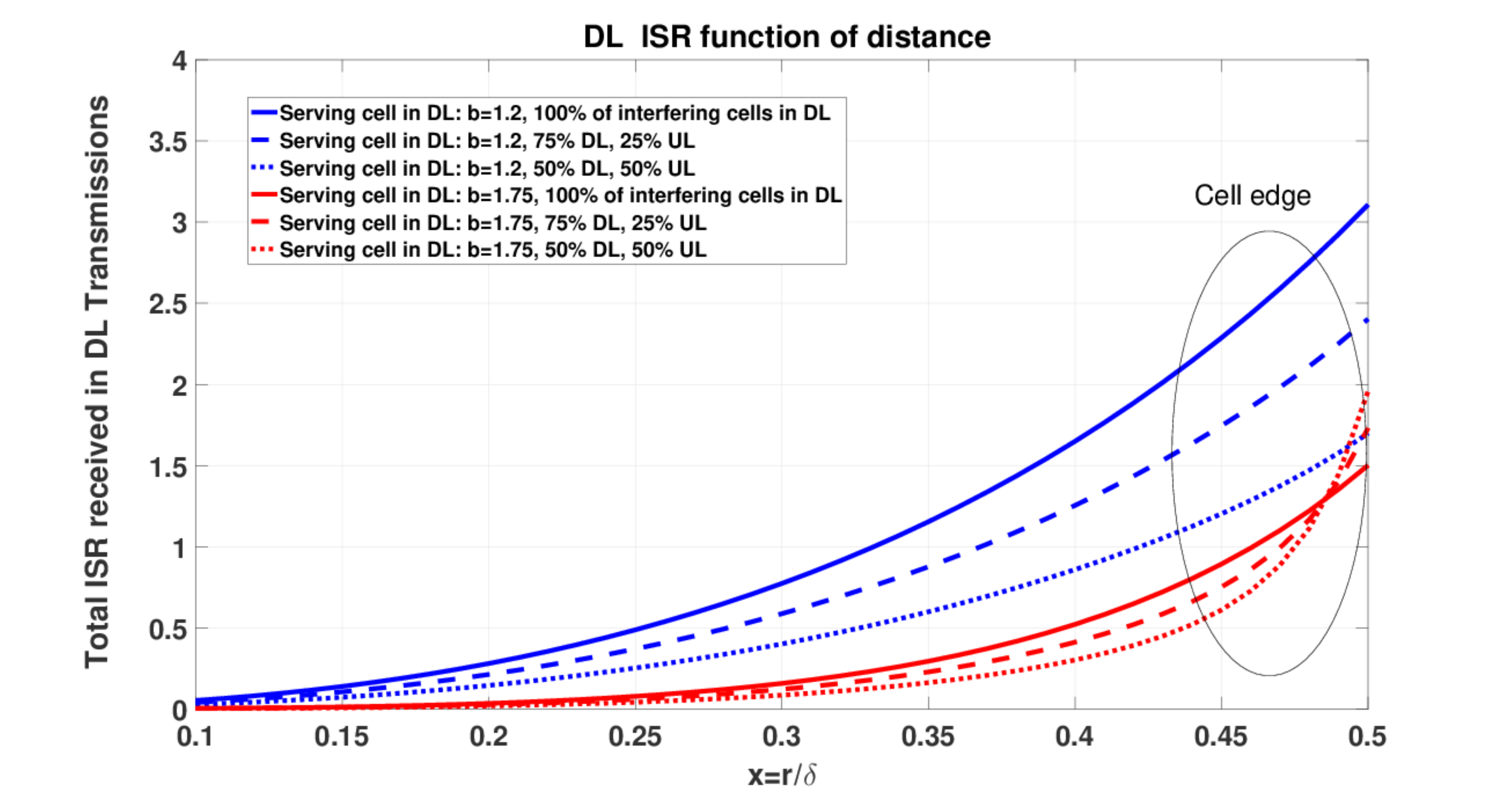}
	\caption{DL ISR in D-TDD system.}
	\label{sim1}
\end{figure}

\begin{figure}[htbp]
	\centering
	\includegraphics[height=5.5cm,width=8cm]{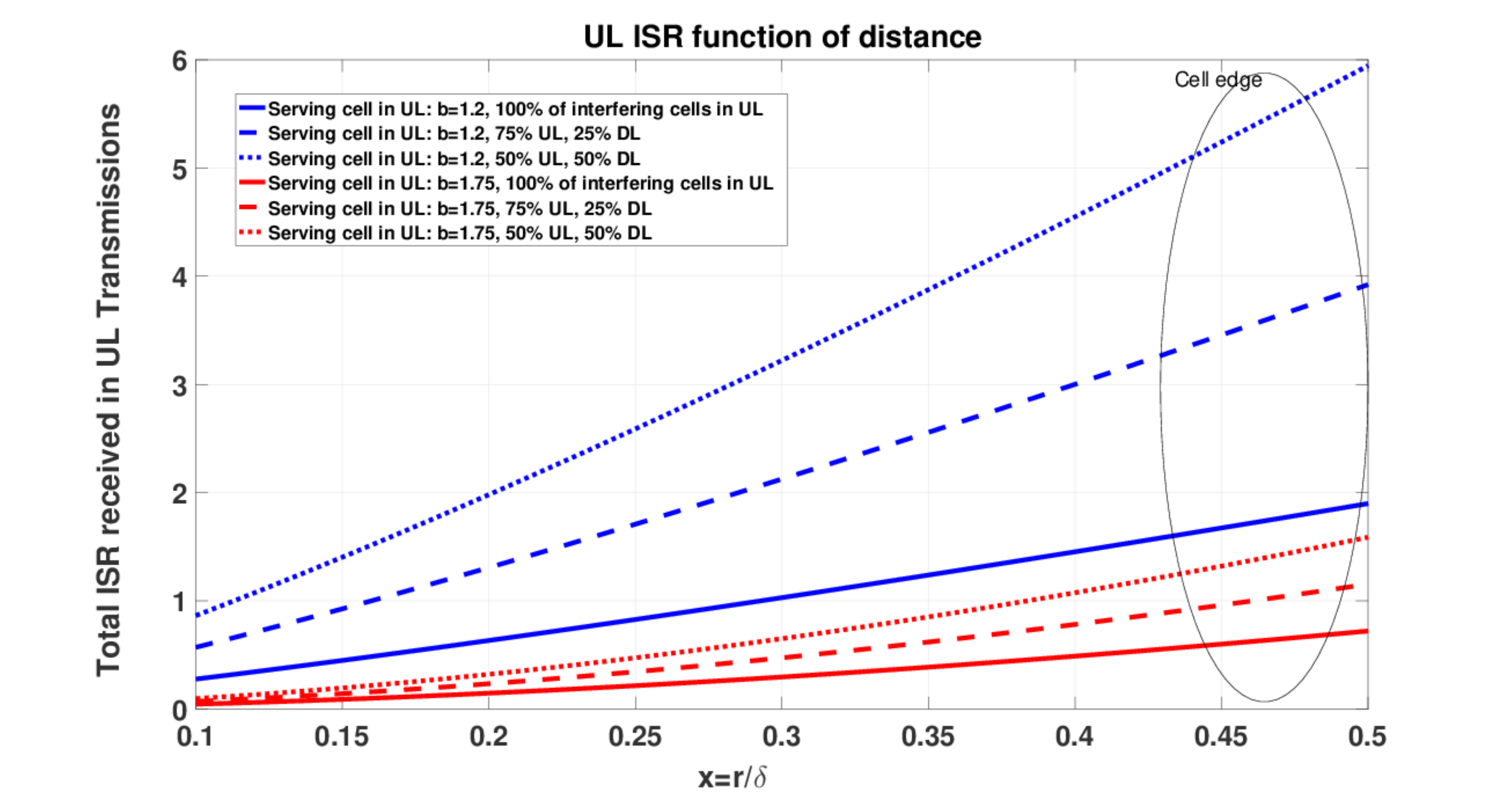}
	\caption{UL ISR in D-TDD system.}
	\label{sim2}
\end{figure}

Fig.\ref{sim1} shows the developed $ISR$ in DL transmission direction for different values of path loss exponent (2b=2.4, 2b=3.5). The first obvious observation is that the DL interference level decreases in the studied cell when other cells use more frequently the UL transmission cycle. This means that the impact of DL interferences coming from other cells is relatively higher than the impact of interferences from mobiles. Consequently, one can conclude that DL interference level in DL cycle for D-TDD should be lower than Static TDD. The system behavior during the UL cycle is completely different. As shown in Fig.\ref{sim2}, interference level significantly increases when 25\% or 50\% of cells switched to the opposite direction, i.e., DL transmission. The UL performance degradation is mainly related to the higher DL transmit power of other cells, especially when they are in LOS conditions. This phenomenon makes D-TDD system very limited by DL to UL interferences. These conclusions are in agreement with the results of  \cite{khoryaev2012performance}, which showed that there is an improvement of 10dB in the DL $SINR$ of the serving cell when 50\% of other cells switch from DL to UL transmission cycle; whereas the UL $SINR$ of the same serving cell degrades by 20dB. This UL performance loss is expected to be more significant in macro-cell deployment. Therefore, DL to UL interference can seriously deteriorate system performances if no action is taken to mitigate it.    

\section{Application: Dynamic TDD interference coordination}

Additional types of interferences that occur in D-TDD systems are the prime concerns to minimize. As mentioned in the introduction, 3GPP advices new approach for interference mitigation in dynamic environment such as cell clustering. Furthermore, HetNets are a good candidate for D-TDD because small cells can be considered well isolated from each others since they transmit with a low power level and the base stations are not highly elevated. In this section we analyze D-TDD performances considering cell clustering feature in heterogeneous deployment.
\begin{figure}
	\centering
	\includegraphics[height=4.5cm,width=7.5cm]{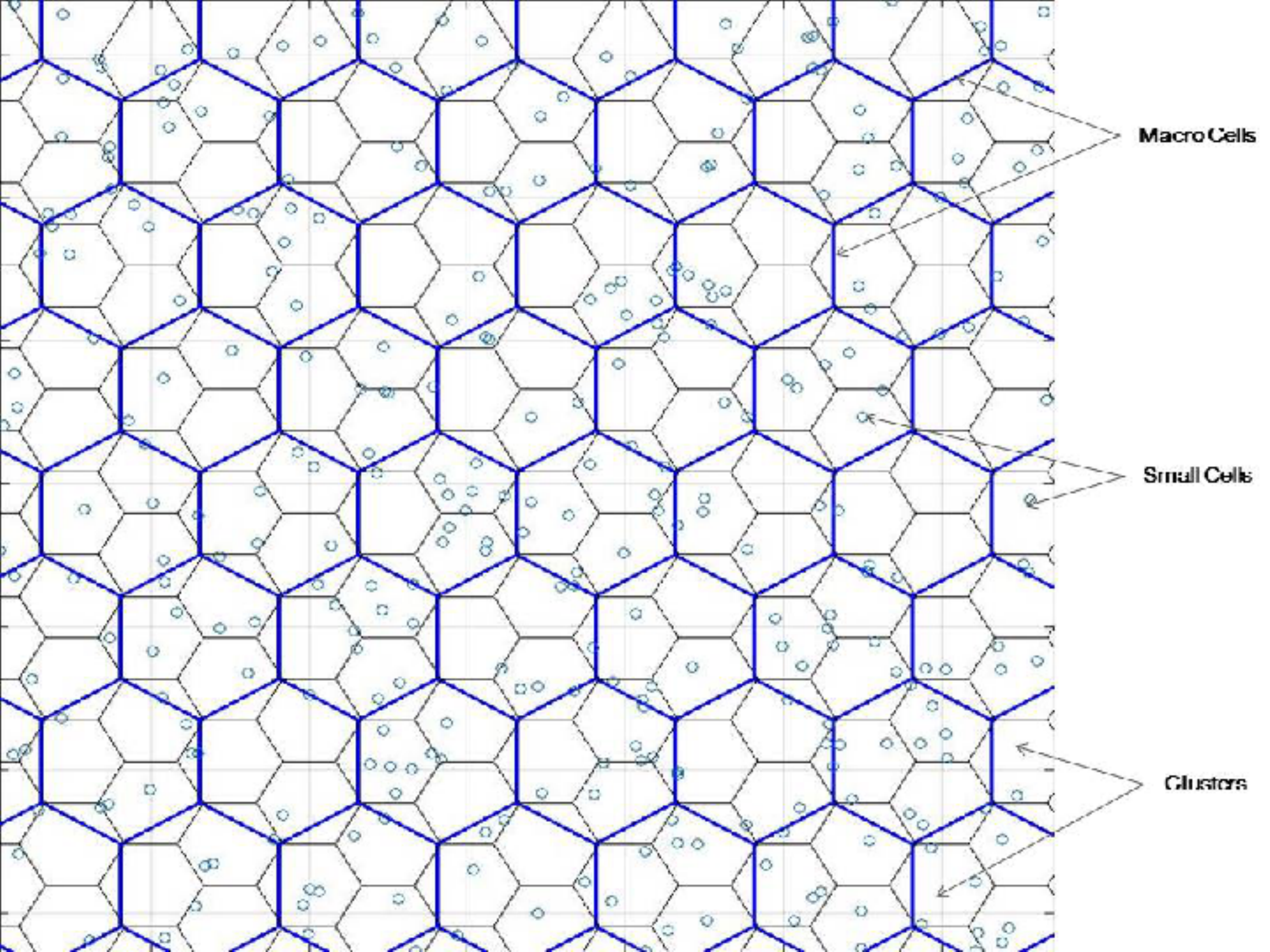}
	\caption{Cell clustering model}
	\label{cluster}
\end{figure} 

The main idea of cell clustering is to gather small cells in different clusters based on specific metrics e.g., Mutual Coupling Loss (MCL) threshold between cells. Small cells in the same cluster adopt the same UL-DL configurations and the transmission inside a cluster can be coordinated by a central unit, which decides the more convenient frame configuration according to traffic conditions. Moreover, cells belonging to different clusters can choose the configuration independently from each others. The problem of cross slot interference still exists even with cell clustering scheme. Actually, the dense deployment of small cells certainly induces severe interference between small cells belonging to neighboring clusters when they transmit in opposite directions. Hence, users served by small cells located closer to the cluster edges can experience bad performances. 

To evaluate this clustering scheme, we consider the same network architecture, propagation and power control models used previously. We assume that clusters form a new hexagonal lattice denoted by $C$ as shown in Fig. \ref{cluster}, with inter-cluster distance $\tilde{\delta}$ such that $\tilde{\delta}= \delta/\sqrt{3}$. Every cluster center $\cluster$ is uniquely identified by the complex variable $\cluster=\tilde{\delta}(u+ve^{i\frac{\pi}{3}})$ with $(u,v) \in \mathbb{Z}^2$ and $\cluster_{0}$ is the new origin of the plane $\mathbb{R}^2$. Cluster radius is denoted by $\tilde{R}$. Each cluster contains $N$ small cells uniformly and independently distributed with the same intensity $\lambda$. Small cell location $\tilde{s}$ is determined by $\tilde{s}= \cluster + \tilde{\rho}e^{i\tilde{\phi}}$ where $\tilde{\rho}$ is the distance between $\tilde{s}$ and the cluster center. The small cell of interest is denoted by $\tilde{\site_{0}}$ such that  $\tilde{\site_{0}}=\tilde{\rho_{0}}e^{i\tilde{\phi_{0}}}$. Small cell radius is designated by $\tilde{R}_{\site}$. A mobile location $\tilde{m}$, served by $\tilde{\site_{0}}$ in UL transmission, is identified by the complex variable $\tilde{m}= \tilde{\site_{0}}+ \tilde{r}e^{i\tilde{\theta}}$, where $\tilde{r}$ is the distance between $\tilde{m}$ and $\tilde{\site_{0}}$. We assume also that small cells transmit with the same power level denoted by $\tilde{P}$. The small cell target power in UL transmission will be denoted $\tilde{P}^{*}$. In the remainder of this section, we assume that macro and small cells operate in different frequency bands, hence interference from macro cells layer will not be considered. Also, UL to UL interference will be neglected and DL cycle will not be analyzed, since cell clustering aims to minimize DL to UL interference in D-TDD system.
\subsection{DL to UL $ISR$ derivation}
Let $ISR(\tilde{\site}_{\downarrow},\tilde{\site_0}_{\uparrow})$ be the individual relative interference received from small cell $\tilde{\site}$ and impacting useful signal at $\tilde{\site_{0}}$ when it is in UL communication with mobile location $\tilde{m}$. Small cell $\tilde{\site}$ is of course assumed to belong to another cluster $\cluster$ of cells operating in DL transmission. The expression of $ISR(\tilde{\site}_{\downarrow},\tilde{\site_0}_{\uparrow})$ can be formulated by
{
\begin{align}
ISR(\tilde{\site}_{\downarrow},\tilde{\site_0}_{\uparrow})=&\frac{\tilde{P}~L(\tilde{m},\tilde{\site_{0}})}{P(\tilde{m},\tilde{\site_{0}})~L(\tilde{\site},\tilde{\site_{0}})}\nonumber\\
=&\frac{\tilde{P} \left|\tilde{\site}-\tilde{\site_0}\right|^{-2b}}{\tilde{P}^{*}~\left|\tilde{m}-\tilde{\site_0}\right|^{2bk}~\left|\tilde{m}-\tilde{\site_0}\right|^{-2b}}\nonumber  
\label{clustindivid}
\end{align}
}
Recalling the fact that small cells are uniformly distributed in clusters with the intensity $\lambda$ (For numerical results, we will take $\lambda=\frac{3}{\pi \tilde{R}^2}$, i.e., three small cells per cluster), DL to UL $ISR$ can be obtained by summing over all small cells belonging to all clusters except the one containing $\tilde{\site_0}$. It follows that
\begin{equation}
U_{\downarrow}(\tilde{m})=\lambda \sum_{c \in C^{*}}\int_{\cluster}ISR(\tilde{\site}_{\downarrow},\tilde{\site_0}_{\uparrow})d\tilde{\site}
\label{isrcluster}
\end{equation}
To evaluate equation (\ref{isrcluster}), we can proceed analogously to the derivation of $D_{\uparrow}$ in section III-A-2. Take $\tilde{x}=\tilde{r}/\tilde{\delta}$, then for $\tilde{x}<1$ and $b>1$, the DL to UL $ISR$ $U_{\downarrow}(m)$ is explicitly written as 
\begin{equation}
U_{\downarrow}(\tilde{m})=\tilde{A}_2(b)~\tilde{x}^{2b(1-k)}
\label{ISRclusteringfinal}
\end{equation}
where 
{
\begin{align}
\tilde{A}_2(b) =& \frac{6 \pi \tilde{R}^2 \tilde{P} \lambda}{\tilde{P}^{*}~\Gamma(b)^{2} \tilde{\delta}^{2bk}}\sum_{h=0}^{+\infty}\sum_{n=0}^{\lfloor \frac{h}{2} \rfloor}\sum_{i=0}^{h-2n} \frac{\Gamma(b+h)^{2} \omega(b+h)}{\Gamma(n+1)^{2}\Gamma(h+1)}\times\nonumber\\
&\frac{ (\frac{\tilde{R}}{\tilde{\rho_{0}}})^{2n+2i}~(\frac{\tilde{\rho_{0}}}{\tilde{\delta}})^{2h}}{\Gamma(i+1)\Gamma(h-2n-i+1) (n+i+1)} \nonumber
\label{psideb}
\end{align}
}

\begin{figure}
	\centering
	\includegraphics[height=5.5cm,width=8cm]{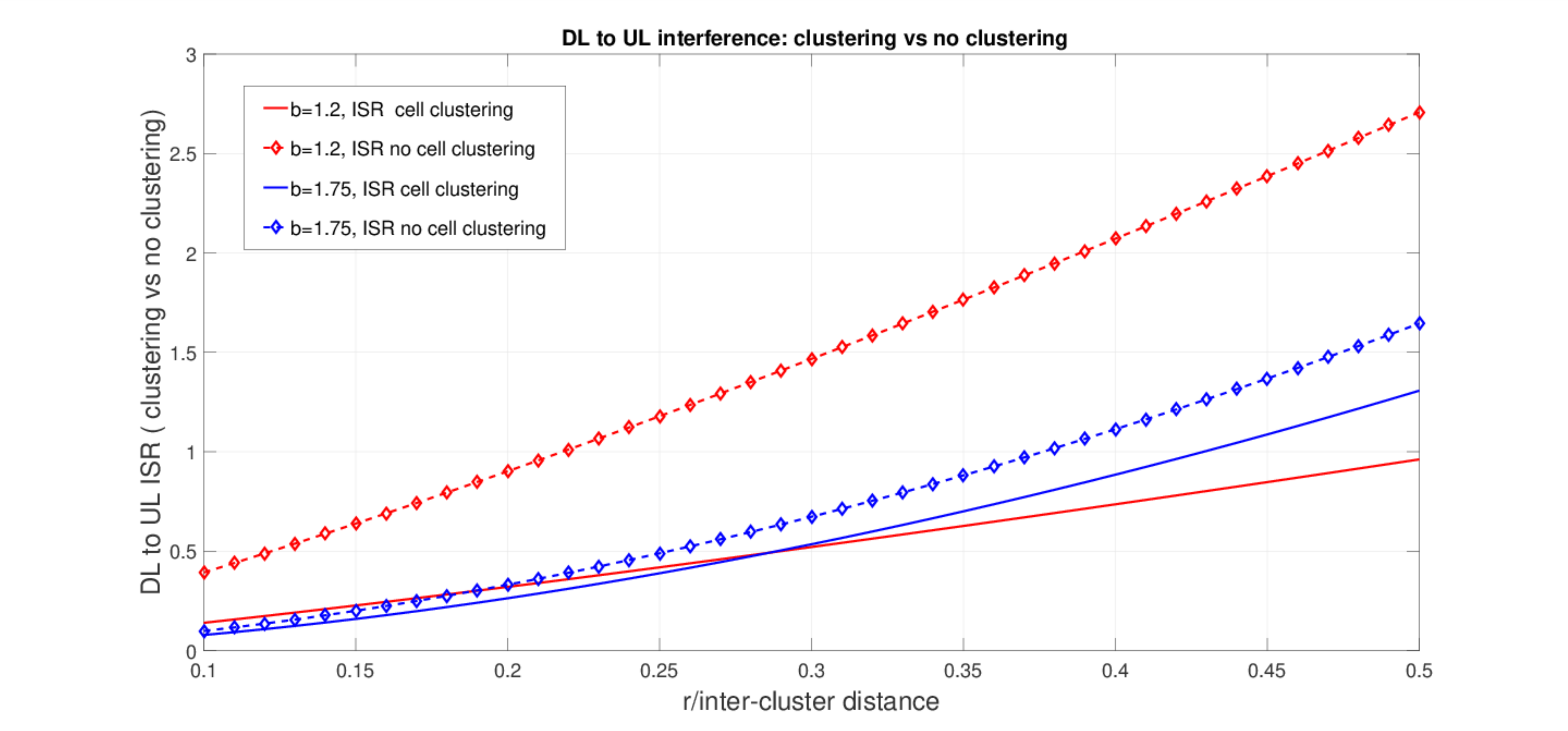}
	\caption{D-TDD DL to UL ISR with cell clustering scheme}
	\label{clustercourbes}
\end{figure}

The strong DL to UL interference, which is the dominant interference in a D-TDD system, comes from the large coupling between small cells. It can be observed from Fig. \ref{clustercourbes} that the clustering scheme minimizes its impact, mainly for favored propagation condition, i.e., low value of parameter $b$.

\subsection{Coverage probability}<
Coverage probability is identified by the percentage of locations having their SINR higher that a given threshold $\gamma$. Let $\Pi(\tilde{x})$ be the UL $SINR$ experienced by small cell $\tilde{\site_0}$ when it communicates with mobile location $\tilde{m}$, such that $|\tilde{m}|=\tilde{x}\tilde{\delta}$.
\begin{equation}
\Pi(\tilde{x})=\frac{1}{U_{\downarrow}(\tilde{x}) + \tilde{y}_{0}\tilde{x}^{2b(1-k)}}
\label{sinrc}
\end{equation}
where $\tilde{y}_{0}=\frac{P_{N}\tilde{\delta}^{2b(1-k)}}{P^{*}}$, with $P_N$ is the noise power. 

Assuming that location $\tilde{m}$ is uniformly distributed in small cell $\tilde{\site}_{0}$, UL coverage probability is evaluated as in \cite{nasri2016Analytical} by:
{
\begin{align}
\Phi(\gamma) =& \mathbb{P}(\Pi(\tilde{x})> \gamma)\nonumber\\
=&\min\left[\left(\frac{\tilde{\delta}}{\tilde{R}_{\site}}g(\frac{1}{\gamma})\right)^2;1\right]
\label{couuuuv}
\end{align}
}
where $g(y)$ is the inverse function of $y(\tilde{x})=1/\Pi(\tilde{x})$. It is given by
\begin{equation}
g(y)=(\frac{y}{\tilde{A}_2(b) + \tilde{y}_{0}})^{\frac{1}{2b(1-k)}}
\end{equation}

\section{Conclusion}
In this paper, we have analytically evaluated inter-cell interferences in D-TDD system. Explicit formulas of $ISR$, covering different interference scenarios in D-TDD, have been derived. Based on the presented analytical results, it can be concluded that D-TDD is only used in favor of DL transmission cycle. However, during an UL transmission, DL to UL interferences may cause a substantial performance degradation. To limit the impact of DL transmissions of other cells on UL transmission, cell clustering scheme is considered by characterizing the experienced DL to UL interferences. It is shown that this feature improves D-TDD systems and somehow reduces the impact of DL to UL transmission. 

\end{document}